# Deep learning based on mixed-variable physics informed neural network for solving fluid dynamics without simulation data

Guang-Tao Zhang[1], Chen Cheng[2], Shu-dong Liu[3], Yang Chen[1], Yong-Zheng Li[3]

1. Department of Mathematics, Faculty of Science and Technology, University of Macau, Macau, China
2. School of Naval Architecture and Ocean Engineering, Jiangsu University of Science and Technology, Jiangsu, China
3. College of Mathematics and Informatics, South China Agricultural University, Guangdong, China

## Abstract

Deep learning method has attracted tremendous attention to handle fluid dynamics in recent years. However, the deep learning method requires much data to guarantee the generalization ability and the data of fluid dynamics are deficient. Recently, physics informed neural network (PINN) is popular to solve the fluid flow problems, which basic concept is to embed the governing equation and continuity equation into loss function, with the requirement of less dataset for obtaining a reliable neural network. In this paper, the mixed-variable PINN method, which convert the governing equation into continuum and constitutive formulations, is proposed to solve the fluid dynamics (flow past cylinder) without any labeled data. The initial/boundary conditions with penalty factors are also embedded into the loss function to become a well-imposed problem. The results show that mixed-variable PINN has better predictive ability to construct the flow field than traditional PINN scheme. Furthermore, the transfer learning method is adopted to is solve the fluid solutions with different Reynold numbers with less computational cost. The results also demonstrate that the transfer learning method can well simulate the different Reynolds number in a short time.

## 1. Introduction

Fluid dynamic problems are ubiquitous in natural and industrial world (e.g. naval architecture, aerospace, civil engineering and automobile dynamics) which has aroused great interest of scholars. Typically, the mathematical models of fluid dynamics are described by the Navier-Stokes (NS) equations, which are high dimensionality and strong nonlinearity of partial differential equations. Computational fluid dynamics (CFD) methods based on numerical simulation, as the mainstream approaches for solving fluid dynamics, has received extensive attention. However, CFD techniques are cumbersome in computational efficiency, especially for solving turbulent flow and complicated geometries. Furthermore, CFD techniques are also limitative in handling the moving mesh and other particular technical means.

Reduced order modeling (ROM) was firstly proposed in optimal design, optimal control and inverse problem application. ROM is a compact pattern of highly fidelity dynamic model. It only retains the most significant components and main effects of model which can reduce the computational efforts and storage spaces. Proper orthogonal decomposition (POD) and dynamic mode decomposition (DMD) are two dominant methods of ROM in solving flow dynamics in lower dimensional representations (Dowell, 1997; Schmid, 2010). Henshawa et al (2007) utilized POD to construct the non-linear model of the aircraft behavior with low dimensionality and evaluate the performance on the real aircraft. Jovanovie et al (2014) developed a sparsity-promoting variant of the standard DMD algorithm to represent the flow field by numerical simulation and then compared to the experiments. The results showed that method can well re-construct the fluid model. Hemati et al (2014) formulated a low-storage approach to perform DMD to simulate the flow past cylinder and compared with the results from particle image velocimetry experiments. However, ROM also has limitations in solving complicated unsteady flows due to the information loss by compressive model.

Deep learning (DL) technology has extraordinary ability to deal with the strong nonlinearity and high dimensionality (LeCun et al, 2015). Recently, DL has a tremendous breakthrough in some fields, such as speech recognition, image processing and event prediction (Goodfellow et al, 2016; Xiong et al, 2015). More recently, DL method is proposed for solving fluid dynamics. Ling et al (2016) constructed the deep learning of RANS turbulence model by embedding Galileo invariant into depth neural network, and firstly realized the prediction of channel flow vortex and separated flow. This is considered to be the first combination of deep neural networks and fluid mechanics (Nathan, 2017). Yeung et al (2017) proposed a deep learning framework for computing Koopman operators of nonlinear dynamic systems, which provides a new idea for modeling nonlinear dynamic systems by combining DMD method with deep neural networks. Miyanawala and Jaiman (2017) predicted the flow characteristics in the wake region of a two-dimensional cylinder by deep convolution network. Jin et al (2018) utilized fusion convolutional neural networks (CNNs) to predict the velocity fields around the circular cylinder by data obtained by pressure fields. Sekar et al (2019) also adopted CNNs technique combined with Multilayer Perceptron (MLP) to calculate the incompressible laminar steady flows. Recurrent neural network (RNN) is another powerful tool to predict temporal features of flow fields. Deng et al (2019) utilized the Long Short-Term Memory (LSTM) to obtain the time coefficient of the flow field. Mohan et al (2019) combined the CNNs and LSTM to predict the spatial-temporal features of turbulence dynamics. However, DL methods require magnanimous data to ensure the prediction accuracy and generalization ability. Furthermore, DL methods build up a surrogate model which is considered as black box and it means that the model lacks physical interpretation.

Raissi et al (2017) firstly proposed physics informed neural network (PINN) to solve the partial differential equations (PDE) and inverse problems. PINN modified the traditional form of the loss function and was embedded with the physical models, with its important breakthrough featuring that the PINN can predict the variables based on physical laws. Tartakovsky et al (2018) utilized PINN to construct the constitutive equations of Decay flow. It demonstrated that PINN has strong performance in solving inverse problems. Moreover, Yang et al (2020) employed Bayesian and PINN to solve the PDE with noisy data. In the aforementioned research, PINN also need a certain amount of training data to achieve the approximate solution of the PDE. As a matter of fact, if the initial conditions and boundary conditions are decided, the unique solutions of the PDE can be predicted by DL without any training data.

The aim of this paper is to propose a mixed-variable PINN method to predict the fluid dynamic problems described by Navier-Stokes equations without any labeled data. A fully-connected neural network (FCNN) is adopted to construct the structure of the DL model, then the conservation equations and the initial/boundary conditions with penalty factors are imposed into the loss function. Compared to the traditional scheme, the N-S equations are converted into the continuum and constitutive formulations. The flow past cylinder is selected to demonstrate to the performance of the proposed method. The performance of proposed approach based on different structures of the FCNN and the different numbers of sampling points are compared and discussed. The structure of the paper can be demonstrated as follow. Section 2 introduces the principle of the FCNN and the back-propagation mechanism. Section 3 describes the loss function embedded with the conservation equations and initial/condition boundaries the optimizer methods are also explained in this section. In Section 4, the mixed-variable PINN method is proposed and how to solve the incompressible flow is introduced. Section 5 compares the performance of proposed approach with the traditional PINN method for solving incompressible flow under different structures of DL and the transfer learning method is selected to calculate the flow problems in other Reynolds numbers. Conclusion is summarized in section 6.

## 2. Network architecture

*2.1 Fully connected neural network*

A deep FCNN structure includes the input layer, hidden layer and output layer. Generally, layer 0 is the input layer while layer $L$ is the output layer, layers $l$ are the hidden layers. Each neuron in hidden layers includes weight, bias and activation function. Generally, activation function plays a significant role in dealing with nonlinear problems. The most frequently used activation functions are sigmoid, tanh and rectified linear units. The structure of the FCNN can be viewed in Fig. 1. The output of a neuron is calculated as follow:

$$z_j^l = w_{jk}^l f_{l-1}(z_k^{l-1}) + b_j^l \tag{1}$$

where $z_j^l$ denotes the output of neuron $j$ in layer $l$; $w_{jk}^l$ the weight between neuron $k$ in layer $l$-1 and neuron j in layer $l$; $f(\cdot)$ the activation function; $b_j^l$ the bias of neuron $j$ in layer $l$. The formula can also be written as:

$$z_j^l = \sum_k w_{jk}^l y_k^{l-1} + b_j^l \tag{2}$$

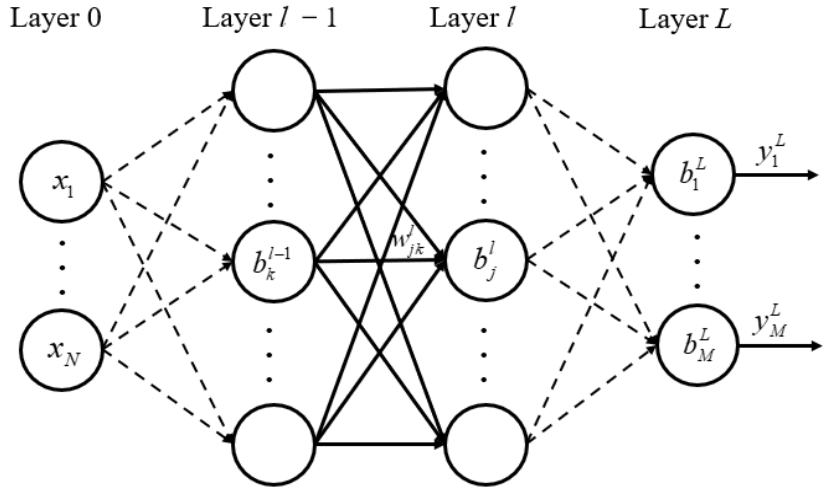

Fig. 1. The structure of the fully-connected neural network

*2.2 Backpropagation*

The input data $x = [x_1, x_2, \cdots x_n]$ and labeled data $y = [y_1, y_2, \cdots y_n]$ are chosen to optimize the neural structure including weights and thresholds. The approximate value $y^L$ predicted by neural network is compared to the real value $y$, the difference between these two values is defined as cost function which can be viewed as follow:

$$w^*, b^* = \arg\min_{w,b} C(y, y^L) \tag{3}$$

where $w^*$ and $b^*$ are optimized weights and thresholds, respectively; $C(y, y^L)$ denotes the cost function. The most significant step is to minimize the cost function to ensure the predicted value is consistent with the real value. Generally, the gradient computation is adopted to accomplish the above process. Backpropagation is a standard approach to compute the gradients and can be viewed as follow:

$$\delta_j^l = \frac{\partial C}{\partial z_j^l} \tag{4}$$

Go a step further, the gradient of the cost function can be computed as another form which can be demonstrated:

$$\delta_j^L = \frac{\partial C}{\partial y_j^L}\sigma_L'\left(z_j^L\right), \qquad \frac{\partial C}{\partial w_{jk}^l} = y_k^{l-1}\delta_j^l,$$

$$\delta_j^l = \sum_k w_{kj}^{l+1}\delta_k^{l+1}\sigma_l'\left(z_j^L\right), \qquad \frac{\partial C}{\partial b_j^l} = \delta_j^l \tag{5}$$

The $\delta$-term in Eq. (5) can be expressed as vectorial form:

$$\delta^L = \nabla_{y_L}C \odot \sigma_L'\left(z^L\right), \qquad \delta^l = \left(W^{l+1}\right)^T \delta^{l+1} \odot \sigma_l'\left(z^l\right) \tag{6}$$

$$\nabla_{y_L}C = \left[\frac{\partial C}{\partial y_1^L},\cdots,\frac{\partial C}{\partial y_M^L}\right]^T \tag{7}$$

where $\odot$ is the Hadamard product. The notation $\nabla C$ without a subscript the vector of partial derivatives in respect of the input $x = [x_1, x_2, \cdots x_n]$.

## 3. FCNN approximations to N-S equations

*3.1 Physics-constrained deep learning*

Conventionally, DL method builds up a surrogate model, such as FCNN or CNN, for predicting the solution of the fluid flow which are approximately equal to real values (Zhu et al, 2018).

$$\mathbf{f}(t,x,\theta) \approx \tilde{\mathbf{f}}(t,x,\theta) \triangleq \mathbf{z}_l(t,x,\theta;\mathbf{W},\mathbf{b}) \tag{8}$$

where $\mathbf{f}$ is the solution vector including the velocity fields and pressure fields; $\mathbf{W}$ and $\mathbf{b}$ denote the weights and biases, respectively. $\mathbf{z}_l(t,x,\theta;\mathbf{W},\mathbf{b})$ the predicted by the surrogate model; $\tilde{\mathbf{f}}$ the locally minimized. The solution of flow dynamics can be cast into an optimization problem which can be demonstrated as follow:

$$L_{data}(\mathbf{W},\mathbf{b}) = \frac{1}{N}\left|\mathbf{f}^d(t,\mathbf{x},\theta) - \mathbf{z}_l(t,\mathbf{x},\theta;\mathbf{W},\mathbf{b})\right|^2$$

$$\mathbf{W}^*,\mathbf{b}^* = \underset{\mathbf{w},\mathbf{b}}{argmin}\, L_{data}(\mathbf{W},\mathbf{b}) \tag{9}$$

where $L_{data}(\mathbf{W},\mathbf{b})$ denotes the loss function based on data; N the number of training samples. $\mathbf{f}^d$ the training data.

However, the traditional DL requires large number of training data, which is too difficult to achieve from time-consuming CFD simulation. Physics-constrained deep learning embeds the physical model into the loss function by minimizing the violation of the solution on the basis of the known partial differential equations for fluid flows over a domain of interests without the demands of handling these equations for each parameter with conventional numerical simulations. The residual of N-S equations and mass conservation equations are computed by FCNN and the specific loss function can be demonstrated as follow:

$$L_{phy}(\mathbf{W},\mathbf{b}) = +\underbrace{\frac{1}{N_f}\sum_{i=1}^{N_f}\left|\frac{\partial \mathbf{u}}{\partial t}+(\mathbf{u}\cdot\nabla)\mathbf{u}+\frac{1}{\rho}\nabla p-\nu\nabla^2\mathbf{u}+\mathbf{b}_f\right|^2}_{Structure\ imposed\ by\ N-S\ equations} + \underbrace{\frac{1}{N_f}\sum_{i=1}^{N_f}\left|\nabla\mathbf{u}\right|^2}_{Mass\ conservation}$$

$$\mathbf{W}^*,\mathbf{b}^* = \underset{\mathbf{w},\mathbf{b}}{argmin}\, L_{phy}(\mathbf{W},\mathbf{b}) \qquad (10)$$

$$s.t.\begin{cases} I\ (\mathbf{x},p,\mathbf{u},\theta)=0, & t=0,\ in\ \Omega_f \\ B\ (t,\mathbf{x},p,\mathbf{u},\theta)=0 & on\ \partial\Omega_{f,t} \end{cases}$$

where $L_{phy}(\mathbf{W},\mathbf{b})$ denotes the physics-based loss; $I$ and $B$ the initial and boundary conditions, respectively;

The first and/or second derivative terms of velocity and pressure in the loss function can be computed by the automatic differentiation approach (AD) (Baydin et al, 2018). Compared to the traditional differential calculation, such as Manual Differentiation, Numerical Differentiation and Symbolic Differentiation, the core problem of AD is to calculate the derivatives, gradients and Hessian matrix values of complex functions, which are usually multi-layer composite functions at a certain point. The advantage of the AD is more accurate due to the absence of truncation or round-off errors. Generally, AD can be directly utilized in deep learning framework such as Tensorflow, Pytorch and Theano (Paszke et al, 2017; Abadi et al, 2016; Bastien et al, 2012). In order to reduce the error of the loss function, the Adam optimizer is utilized to optimize the target function. Adam optimizer can constantly adjust the learning rates with the situation changes in the learning process (Diederik and Jimmy, 2017). 'Xavier' method is designed to decide the initial weights and biases which can ensure faster convergence of neural network (Glorot and Bengio, 2010). A residual neural network is added in the FCNN to avoid gradient explosion and/or gradient disappearance (He et al, 2016).

*3.2 Initial and Boundary condition enforcement*

The loss function constrained by the physical equations becomes identically zero, the predicted values of velocity and pressure fields will precisely satisfy the N-S equations. Consequently, the solutions driven by FCNN particularly have physical interpretation through penalizing the PDE residuals. Furthermore, to make the problem well-posed, the appropriate initial conditions and boundary conditions are required and imposed as constraints which are dealt with a soft manner by amending the original loss function with penalty terms (Márquez-Neila et al, 2017). The Eq. (10) can be rewritten by adding initial loss and boundary loss as follow:

$$L_{phy}^{c}(\mathbf{W},\mathbf{b},\lambda_i,\lambda_b) = \underbrace{L_{phy}(\mathbf{W},\mathbf{b})}_{Equation\ loss} + \underbrace{\lambda_i\left\|I\ (\mathbf{x},p,\mathbf{u},\theta)\right\|_{\Omega_{f,t}}}_{Initial\ loss} + \underbrace{\lambda_b\left\|B\ (t,\mathbf{x},p,\mathbf{u},\theta)\right\|_{\partial\Omega_{f,t}}}_{Boundary\ loss} \qquad (11)$$

where $\lambda_i$ and $\lambda_b$ are penalty coefficients.

## 4. Mixed-variable PINN for solving incompressible flow

The incompressible flow past cylinder is a classic fluid-structure interaction problem, it can be described by continuity equation and N-S equations as follow:

$$\nabla \cdot \mathbf{v} = 0 \tag{14}$$

$$\frac{\partial \mathbf{v}}{\partial t} + (\mathbf{v} \cdot \nabla)\mathbf{v} = -\frac{1}{\rho}\nabla p + \frac{\mu}{\rho}\nabla^2 \mathbf{v} + \mathbf{b}_f \tag{15}$$

where $\nabla$ is the Nabla operator, $\mathbf{v}$ and $\mathbf{p}$ are velocity field and pressure field, respectively. $\mu$ denotes the viscosity of the fluid, $\rho$ the density of fluid and $\mathbf{b}_f$ the body force. Typically, the aforementioned partial differential equations are embedded into the loss function to calculate the gradients. However, the predicted results are intractable to obtain due to its complicated form of latent variable (such as $\mathbf{v}$ and $p$) and high-order derivatives (such as $\nabla^2$). To make the loss functions easier to be trained, the N-S equations are transformed to the following continuum and constitutive formulations:

$$\frac{\partial \mathbf{v}}{\partial t} + (\mathbf{v} \cdot \nabla)\mathbf{v} = -\frac{1}{\rho}\nabla \cdot \sigma + \mathbf{b}_f \tag{16}$$

$$\sigma = -p\mathbf{I} + \mu(\nabla \mathbf{v} + \nabla \mathbf{v}^T) \tag{17}$$

where $\sigma$ denotes the Cauchy stress tensor and can also expressed as tensor form:

$$\begin{bmatrix} \sigma_{11} & \sigma_{12} \\ \sigma_{21} & \sigma_{22} \end{bmatrix} = \begin{bmatrix} -p & 0 \\ 0 & -p \end{bmatrix} \begin{bmatrix} 1 & 0 \\ 0 & 1 \end{bmatrix} + \mu \begin{bmatrix} 0 & \tau_{12} \\ \tau_{21} & 0 \end{bmatrix} \tag{18}$$

$$\frac{\partial \sigma_{ij}}{\partial x_j} = \frac{\partial}{\partial x_j}\left[-p\delta_{ij} + \mu\left(\frac{\partial u_i}{\partial x_j} + \frac{\partial u_j}{\partial x_i}\right)\right] \tag{19}$$

In addition, every second-order tensor can be decomposed into deviatoric and hydrostatic parts which can be demonstrated as follow:

$$\sigma = \sigma^{dev} + \sigma^{hyd} \tag{20}$$

The hydrostatic part of any Cauchy stress matrix has negative pressure and can be expressed:

$$-p = \sigma^{hyd} = \frac{1}{2}tr(\sigma) \tag{21}$$

In this section, the proposed DL constrained by physical laws to model transient flows past cylinder. The constant velocity condition is imposed on the inlet while the zero-pressure condition is enforced on the outlet, as demonstrated in Fig. 2. Non-slip conditions are applied on the wall and cylinder surfaces. In addition, the body force is ignored in this case. The density and viscosity of fluid are 1 $kg/m^3$ and 0.01 $kg/(m*s)$, respectively. The structure of the fully-connected neural network for solving N-S equations can be viewed in Fig. 3. The spatiotemporal variables $\{t, \mathbf{x}\}$ and mix-variable results $\{\psi, p, \sigma\}$ build up the DNN structure. It is noteworthy that the steam function $\psi$ is adopted instead of the velocity field $\mathbf{v}$ to guarantee the conditions of divergence free flow. By this means, the continuity formulation can also be met automatically. Therefore, the loss function embedded by the physical laws can be formulated as follow:

$$L_{phy}(\mathbf{W},\mathbf{b}) = \underbrace{\left\|\frac{\partial \mathbf{v}}{\partial t} + (\mathbf{v}\cdot\nabla)\mathbf{v} + \frac{1}{\rho}\nabla\cdot\sigma\right\|}_{Continuum\ equation} + \underbrace{\left\|\sigma + p\mathbf{I} - \mu(\nabla\mathbf{v} + \nabla\mathbf{v}^T)\right\|}_{Constitutive\ equation} + \underbrace{\left\|p + \frac{1}{2}tr(\sigma)\right\|}_{Stress-pressure\ equation} \quad (22)$$

Furthermore, user-defined penalty coefficients are introduced to handle the initial conditions and boundary conditions and the total loss function can be described:

$$L_{phy}^c(\mathbf{W},\mathbf{b},\lambda_i,\lambda_b) = \underbrace{L_{phy}(\mathbf{W},\mathbf{b})}_{Equation\ loss} + \underbrace{\lambda_i\left(\|\mathbf{u}(x,y,0) - \mathbf{u}_0\| + \|p(x,y,0) - p_0\|\right)}_{Initial\ loss}$$
$$+ \underbrace{\lambda_b\left(\underbrace{\|\mathbf{u} - l\|}_{Dirichlet} + \underbrace{\|\sigma\cdot n - h\|}_{Neumann}\right)}_{Boundary\ loss} \quad (23)$$

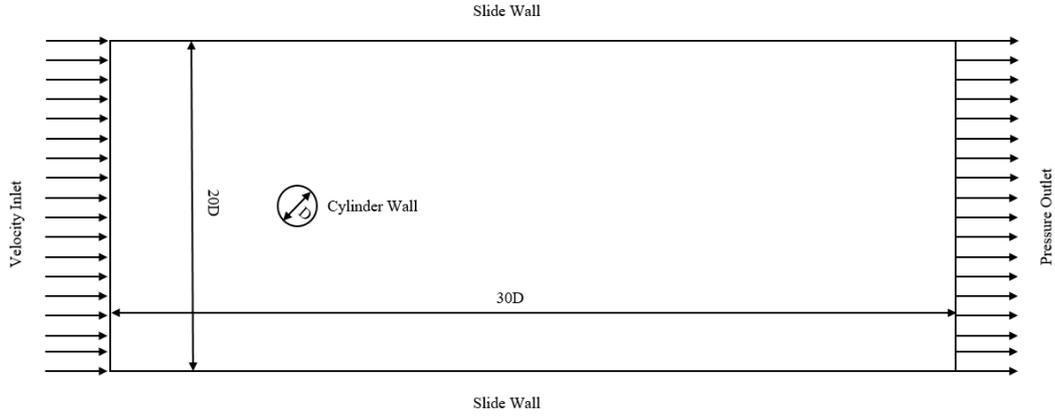

Fig. 2. The scenario of the incompressible flow around the cylinder

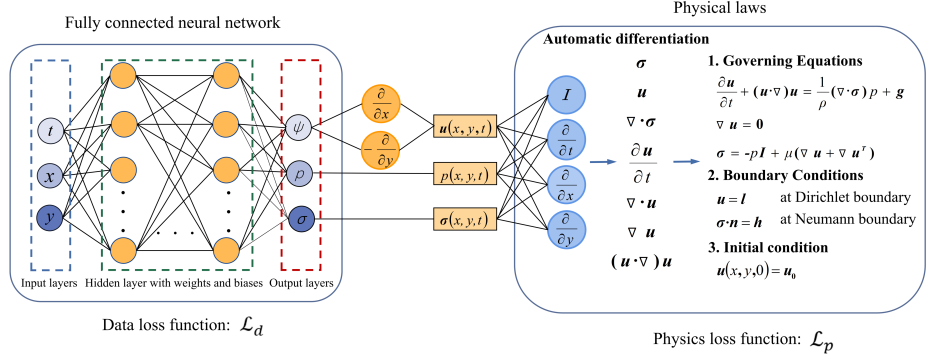

Fig. 3. The structure of the fully-connected neural network for solving N-S equations

Since the training samples are absent, the collocation points in the spatiotemporal zone are required to calculate the loss function. The Latin hypercube sampling (LHS) method is adopted to generate the points in Dirichlet boundary (cylinder, wall, inlet), Neumann boundary (outlet) and other parts in computational zone. It is noteworthy that the collocation points are refined (more collocation points) around the cylinder to better capture the characteristics of the flow. The collocation points in the computational zone can be described in Fig. 4. The green collocation points represent the wall and cylinder surface, the red collocation points and orange collocation points denote the inlet and outlet, respectively. A grid search strategy is selected to obtain an optimal neural network and relative $L_2$-norm error is utilized to judge the accuracy of the calculation which can be demonstrated as:

$$\varepsilon = \frac{\sqrt{\sum_{i=1}^{N}\left|f_{pred}^{i} - f_{ref}^{i}\right|^{2}}}{\sqrt{\sum_{i=1}^{N} f_{ref}^{i}}} \qquad (24)$$

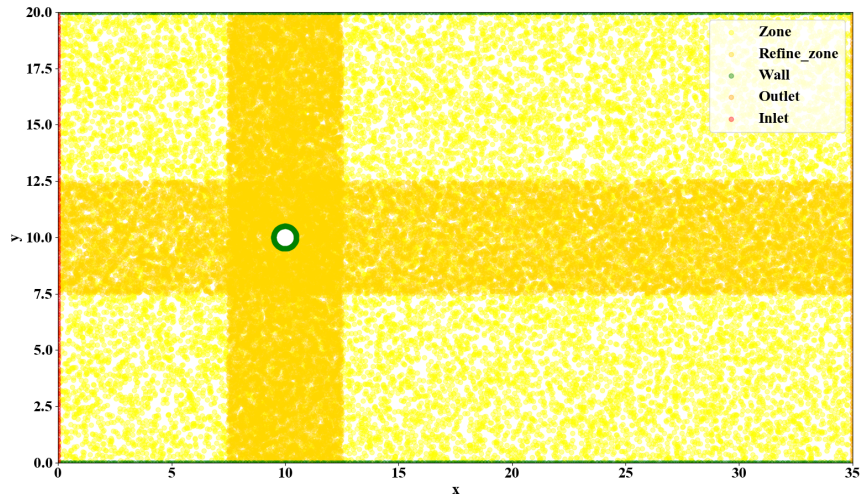

Fig. 4. The Collocation points in the computational zone

## 5. Discussion

*5.1 Physical-constrained learning for flow past cylinder*

In this section, the flow past cylinder in low Reynolds number is adopted to calculate. The inlet velocity is 1m/s and the Reynolds number is 100. The Adam and Limited-memory BFGS optimizer is selected to optimize the weights and thresholds in DL structure. The traditional scheme (Eq. 1 and Eq. 2) is also selected, where the steam function and pressure are adopted as outputs, to compare the proposed model in this paper. The different structures of the neural network are compared and the mean square errors under traditional scheme and continuum and constitutive formulations which can be viewed in Table 1. It can be observed that the structure of the neural networks including 7 layers with 60 neurons has the best capacity to calculate the field information. In addition, the mix-variable deep learning method has the better predictive ability than the traditional PINN.

The velocity field and pressure field predicted by the mixed-variable PINN can be viewed in Fig. 5. The exact solution is obtained from the Openfoam 5.x package based on finite volume method. It can be demonstrated that the mixed-variable PINN can well simulate the velocity field and pressure field. It is interesting to note that the pressure value around the circular is concerned and the mean square error of this part is $3.25 \times 10^{-3}$. It means that the drag force and lift force can be well predicted by calculate following equations (show in Fig. 6):

$$F_L = \int_\Gamma \left[ -pn_y + \frac{2}{Re}\frac{\partial v}{\partial y}n_y + \frac{1}{Re}\left(\frac{\partial u}{\partial y}+\frac{\partial v}{\partial x}\right)n_x \right]ds$$
$$F_D = \int_\Gamma \left[ -pn_x + \frac{2}{Re}\frac{\partial u}{\partial x}n_x + \frac{1}{Re}\left(\frac{\partial u}{\partial y}+\frac{\partial v}{\partial x}\right)n_y \right]ds \quad (26)$$

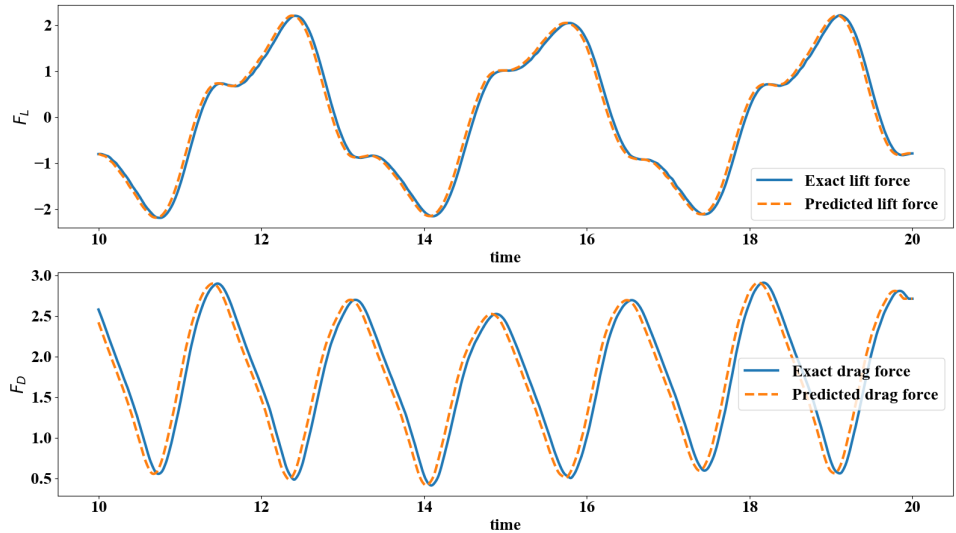

Fig. 6 Lift force and drag force are predicted by mixed-variable PINN

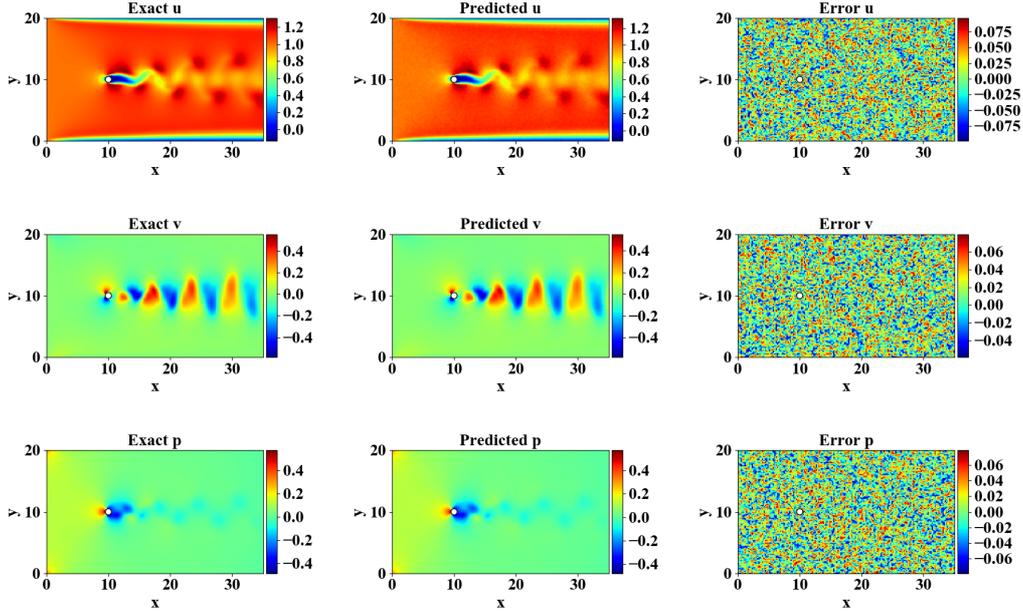

Fig. 5. Performance of the proposed DNN for solving a N-S equation with low Reynolds number

Table 1. Mean square errors of the velocity field under different structures of neural network (left is the traditional scheme, right is proposed scheme)

| Width depth | 5 | 6 | 7 | 8 |
|---|---|---|---|---|
| 20 | $8.64\times10^{-1}$/ $1.32\times10^{-1}$ | $7.52\times10^{-1}$/ $0.82\times10^{-1}$ | $7.82\times10^{-1}$/ $0.61\times10^{-1}$ | $7.14\times10^{-1}$/ $0.27\times10^{-1}$ |
| 30 | $6.42\times10^{-1}$/ $1.13\times10^{-1}$ | $6.14\times10^{-1}$/ $1.15\times10^{-1}$ | $5.89\times10^{-1}$/ $0.42\times10^{-1}$ | $4.91\times10^{-1}$/ $0.21\times10^{-1}$ |
| 40 | $5.99\times10^{-1}$/ $0.86\times10^{-1}$ | $6.14\times10^{-1}$/ $0.54\times10^{-1}$ | $5.41\times10^{-1}$/ $0.42\times10^{-1}$ | $5.13\times10^{-1}$/ $0.45\times10^{-1}$ |
| 50 | $5.14\times10^{-1}$/ $0.71\times10^{-1}$ | $5.88\times10^{-1}$/ $0.51\times10^{-1}$ | $5.53\times10^{-1}$/ $0.26\times10^{-1}$ | $5.49\times10^{-1}$/ $0.75\times10^{-2}$ |
| 60 | $5.02\times10^{-1}$/ $0.54\times10^{-2}$ | $4.35\times10^{-1}$/ $0.27\times10^{-2}$ | $4.87\times10^{-1}$/ $0.89\times10^{-3}$ | $4.14\times10^{-1}$/ $0.92\times10^{-3}$ |

*5.2 Transfer learning of mixed-variable PINN in other Reynolds number*

Previous section focus on the laminar flow at low Reynolds number and shows strong predictive ability to construct the whole flow field. Therefore, the mixed-variable PINN is in theory applicable to higher Reynolds or even turbulent flows. However, it means that the much finer collocation points are required in the computational zone which will result in increasing the computational expense.

In this section, the transfer learning method is adopted to handle the above problem. The definition of transfer learning is to recognize and apply knowledge and skills learned in previous domains/tasks to novel domains/tasks. The parameters of the well-trained mixed-variable PINN in a small Reynolds number can be transferred to mixed-variable PINN in a larger Reynolds, thus fewer iterations are needed to obtain the accurate velocity and pressure. We have already obtained the mix-variable PINN for $Re = 100$ with exact solutions, we want to obtain the solution of velocity and pressure at $Re = 200, 300, 500, 800, 1000$. Firstly, the weights and thresholds are initialized by the pre-trained mix-variable PINN for $Re = 100$, then the weights and thresholds in mix-variable are fine-tuned based on new boundary conditions or governing equations. In the process of transfer learning, only L-BFGS-B is utilized to train the model due to the initialization is close to exact values.

The mean square error of velocity/pressure values and computational expense for mix-variable PINN and transfer learning are demonstrated in Table 2. The results show that the transfer learning method can well predict the velocity and pressure in flow past cylinder under higher Reynolds number and not increase much computer cost due to the lower iterations. Therefore, the mix-variable PINN method can well construct the whole fluid field without any simulation data for different Reynolds numbers with less computational expense.

Table 2. Mean square errors and computational cost of the velocity field and pressure under different Reynolds by transfer learning method

|  | $\varepsilon_u$ | $\varepsilon_v$ | $\varepsilon_p$ | Computer cost(min) |
|---|---|---|---|---|
| $Re = 100$ | $0.92 \times 10^{-3}$ | $0.85 \times 10^{-3}$ | $3.25 \times 10^{-3}$ | 207.7 |
| $Re = 200$ | $0.89 \times 10^{-3}$ | $1.32 \times 10^{-3}$ | $2.74 \times 10^{-3}$ | 226.9 |
| $Re = 300$ | $1.19 \times 10^{-3}$ | $2.21 \times 10^{-3}$ | $3.13 \times 10^{-3}$ | 232.5 |
| $Re = 500$ | $1.52 \times 10^{-3}$ | $1.85 \times 10^{-3}$ | $4.16 \times 10^{-3}$ | 218.8 |
| $Re = 800$ | $1.35 \times 10^{-3}$ | $2.96 \times 10^{-3}$ | $3.57 \times 10^{-3}$ | 244.4 |
| $Re = 1000$ | $1.82 \times 10^{-3}$ | $3.11 \times 10^{-3}$ | $3.89 \times 10^{-3}$ | 251.6 |

## 6. Conclusion

In this section, the mixed-variable PINN is proposed to solve the fluid flows without any simulation data. The governing equation and initial/boundary conditions with penalty factors are embedded into the loss function. The fully-connected neural network is selected to construct the structure of the deep learning. The Adam and L-BFGS-B optimizers are adopted to optimize the loss function. The Latin hypercube sampling (LHS) method is adopted to generate the points in computational zone. The flow past cylinder are concerned to test the proposed method in this paper. Compared to the traditional PINN, the governing N-S equations are converted into the continuum and constitutive formulations, called mix-variable PINN, that construct the IC/BC-encoded physics-constrained deep learning. The results show that proposed method has better solving ability to obtain the velocity and pressure than traditional PINN scheme. Furthermore, transfer learning method is utilized to solve the fluid solutions with different Reynold numbers with less computational cost.

**Reference**

Dowell E. H. 1997. Eigen mode analysis in unsteady aerodynamics: Reduced-order models. *Applied Mechanics Review*, 50(6), 371.

Schmid P. J. 2010. Dynamic mode decomposition of numerical experimental data. *Journal of Fluid Mechanics*, 656(10), 5-28.

Henshawa M. J., Badcock K. J., Vio G. A. 2007. Non-linear aeroelastic prediction for aircraft applications. *Progress in Aerospace Science*, 43(4-6), 65-137.

Jovanovie M. R., Schmid P. J., Nichols J. W. 2014. Sparsity-promoting dynamic mode decomposition. *Physics of Fluids*, 26(2), 024103.

Hemati M. S., Williams M. O., Rowley C. W. 2014. Dynamic mode decomposition for large and steaming datasets. *Physics of Fluids*, 26(11), 111701.

LeCun Y., Bengio Y., Courville A. 2015. Deep learning. *Nature*. 521(7553), 436-444.

Goodfellow I., Bengio Y., Courville A. Deep learning (*MIT press*, 2016).

Sainath T. N., Mohamed A., Kingsbury B., Ramabhadran B. Deep convolutional neural networks for LVCSR, in *2013 IEEE International Conference on Acoustics, Speech and Signal Processing (ICASSP)* (IEEE, 2013), 8614–8618.

Ling J., Andrew, K. 2016. Reynolds averaged turbulence modelling using deep neural networks with embedded invariance. *Journal of Fluid Mechanics*, 807: 155-166.

Nathan K. J. 2017. Deep learning in fluid dynamics. *Journal of Fluid Mechanics*, 814: 1-4.

Yeung E., Kundu S., Hodas N. 2017. Learning deep neural network representations for Koopman operators of nonlinear dynamical systems. arXiv:1708.06850v2.

Miyanawala T. P., Jaiman R. K. 2017. An efficient deep learning technique for the Navier-Stokes Equations: application to unsteady wake flow dynamics. arXiv:1710.09099

Jin X., Cheng P., Chen W., Li H. 2018. Prediction model of velocity field around circular cylinder over various Reynolds numbers by fusion convolutional neural networks based on pressure on the cylinder. *Physics of Fluids*, 30, 047105.

Sekar V., Jiang Q., Shu C., Khoo B. 2019. Fast flow field prediction over airfoils using deep learning approach. *Physics of Fluids*, 31, 057103.

Deng Z., Chen Y., Liu Y., Kim K. 2019. Time-resolved turbulent velocity field reconstruction using a long short-term memory (LSTM)-based artificial intelligence framework. *Physics of Fluids*, 31, 075108.

Mohan A. T., Daniel D., Chertkov M., Livescu D. 2019. Compressed convolutional LSTM: An efficient deep learning framework to model high fidelity 3D turbulence. arXiv:1903.00033.

Raissi M., Perdikaris P., 2017. Physics-informed neural networks: A deep learning framework for solving forward and inverse problems involving nonlinear partial differential equations. arXiv:1808.03398.

Tartakovsky A.M., Marrero C.O., Tartakovsky D., Barajas-Solano D. Learning parameters and constitutive relationships with physics informed deep neural networks. arXiv:1808.03398

Yang L., Meng X. H., Karniadakis G. 2020. B-PINN: Bayesian physics-informed neural networks for forward and inverse PDE problems with noisy data. *Journal of Computational Physics*, 425:1-23.

Zhu Y., Zabaras N. 2018. Bayesian deep convolutional encoder-decoder networks for surrogate modeling and uncertainty quantification. *Journal of Computational Physics*, 375: 415-447.

Baydin A. G., Pearlmutter B. A., Radul A. A., Siskind J. M. 2018. Automatic differentiation in machine learning: a survey. Journal of Machine Learning Research, 18: 1-43.

Paszke A., Gross S., Chintala S., Chanan G., Yang E., DeVito Z., Lin Z., Desmaison A., Antiga L., Lerer A. Automatic differentiation in PyTorch, in: NIPS Autodiff Workshop, 2017.

Abadi M., Barham P., Chen J., Chen Z., Davis A., Dean J., Ghemawat S., Irving G. et al. Tensorflow: A system for large-scale machine learning, in: 12th USENIX Symposium on Operating Systems Design and Implementation, OSDI 16, 2016, 265–283.

Bastien F., Lamblin P., Pascanu R., Bergstra J., Goodfellow I. J., Bergeron A.,

Bounchard N., Bengio N. Theano: new features and speed improvements. deep learning and unsupervised feature learning, in: Neural Information Processing Systems Workshop (NIPS), 2012, 1–10.

Diederik P. K., Jimmy L. B. Adam: A Method for Stochastic Optimization. 2017. arXiv:1412.6980v9

Glorot X., Bengio Y. 2010. Understanding the difficulty of training deep feedforward neural networks. Journal of Machine Learning, 249-256.

He K., Zhang X., Ren S., et al. 2016.Deep residual learning for image recognition. Proceedings of the IEEE conference on computer vision and pattern recognition, 770-778.

Márquez-Neila P., Salzmann M., Fua P. Imposing hard constraints on deep network: Promises and limitations. arXiv: 1706.02025.

**Acknowledgement**
This work was supported by the Fundamental Research Fund for the Central Universities of China, and the Postgraduate Research (B200203073) and Practice Innovation Program of Jiangsu Province (KYCX20_0483).